\documentclass[reprint,aps]{revtex4-2}
\usepackage{graphicx}
\usepackage{dcolumn}
\usepackage{bm}
\usepackage{amsmath,amssymb,amsthm,mathrsfs,amsfonts,dsfont}
\usepackage{bbold}
\usepackage{float}
\newcommand{\id}{\mathbb{1}}

\newcommand{\HS}{\mathcal{H}}
\newcommand{\N}{\mathcal{N}}
\newcommand{\LI}{\mathcal{L}}

\newcommand{\C}{\mathcal{C}}
\newcommand{\dagg}{\dagger}

\newcommand{\oprod}[2]{| #1 \rangle\!\langle #2 |}

\usepackage{braket}

\begin{document}

\preprint{APS/123-QED}

\title{Channel capacity enhancement with indefinite causal order}

\author{Nicolas Loizeau}
\affiliation{Department of Physics, New York University, 726 Broadway, New York, NY 10003, USA}

\author{Alexei Grinbaum}%
\affiliation{%
CEA-Saclay, IRFU/Larsim, 91191 Gif-sur-Yvette Cedex, France
}%

\date{\today}

\begin{abstract}
Classical communication capacity of a channel can be enhanced either through a device called a `quantum switch' or by putting the channel in a quantum superposition. The gains in the two cases, although different, have their origin in the use of a quantum resource, but is it the same resource?
Here this question is explored through simulating large sets of random channels. We find that quantum superposition always provides an advantage, while the quantum switch does not: it can either increase or decrease communication capacity. The origin of this discrepancy can be attributed to a subtle combination of superposition and non-commutativity.
\end{abstract}

\maketitle

\section{Introduction}

In spacetime, events $A$ and $B$ can be in three causal relations: either $A$ is before $B$, $B$ is before $A$, or $A$ and $B$ are causally separated, i.e. they lie on a spacelike interval. Quantum mechanics admits causal structures that do not correspond to any of these cases. Heuristically, this can be pictured as putting the order between $A$ and $B$ in a quantum superposition. More precisely, several approaches to indefinite causal orders have been proposed using `process matrix' or `quantum switch' \cite{oreshkov_quantum_2012,chiribella_quantum_2013,hardy_towards_2007,Brukner2015_2,Chiribella2019,Issam1}. While these approaches are not strictly equivalent mathematically, all of them support one underlying idea: an indefinite causal order is an inherently quantum phenomenon that sheds new light on a notion hitherto explored mainly in spacetime theories. This phenomenon has recently been observed experimentally in several implementations of the quantum switch~\cite{BruknerWalther,Rubinoe,goswami1,goswami2,PhysRevLett.122.120504,Procopio2019}.

To gauge precisely the new element brought by quantum theory into the study of causality, quantum control on causal order can be treated as a resource that provides non-classical communication advantage, i.e., two noisy channels in a quantum switch can transmit more information than any of these channels individually~\cite{Ebler}. This approach has the benefit of immediately clarifying the physical interest of the quantum switch, however it
relies on a currently unresolved question whether any local party can operationally exercise such quantum control~\cite{Oreshkov2018}. In this work we assume that a positive heuristic has been given by the empirical work: quantum control of the causal order via the quantum switch has been obtained experimentally. In what follows we strive to achieve a better theoretical understanding of the advantage demonstrated in such setups. In particular, a standing problem concerns the origin of this advantage: to deny that the quantum switch is an independent resource, it has been argued that the one-pass quantum superposition of two channels, without the indefinite causal order, already leads to a similar result \cite{abbott,Brukner2019}.

After introducing basic mathematical concepts in Section \ref{sect:math}, we explore the controversial origin of this non-classical advantage in Section \ref{section:results}. To this end, we simulate large sets of random channels and use them to compare communication gains from the quantum switch and the one-pass superposition. In Section \ref{section:discussion}, we argue that the advantage for the quantum switch has its origin in two separate factors. One is quantum superposition; the other is non-commutativity of the Kraus decompositions of the channels. A combination of these factors can be significantly more beneficial than the advantage gained from the superposition alone, but in other cases it can also be much less advantageous. When the indefinite causal order is realized through a quantum switch, the gain provided by this resource is essentially due to this combination.

\section{Mathematical framework}
\label{sect:math}
A quantum system going through a quantum channel is modelled by a completely positive trace preserving linear map on its state Hilbert space $\HS$. Any such map $\C$ can be represented by a set of Kraus operators $\{K_i\}\subset \LI (\HS)$ such as \citep{CHOI1975,nielsenchuang,Kraus}:
\begin{equation}
\C(\rho)=\sum_i K_i \rho  K_i^\dagger \quad \text{and} \quad \sum_i K_i^\dagger K_i = \mathbb{1}.\label{kraus}
\end{equation}
This decomposition is not unique: if $\C$ and $\C^\prime$ have Kraus operators $\{K_i\}$ and $\{ K^\prime _i \}$ respectively, then $\C$ implements the same channel as $\C^\prime$ if and only if there exists a unitary operator $u$ such as:
\begin{equation}
    K_{i} = \sum_j u_{ij} K^\prime_j
    \label{kraussfree}.
\end{equation}

\begin{figure}[!htb]
\centering
\includegraphics[width=0.4\textwidth]{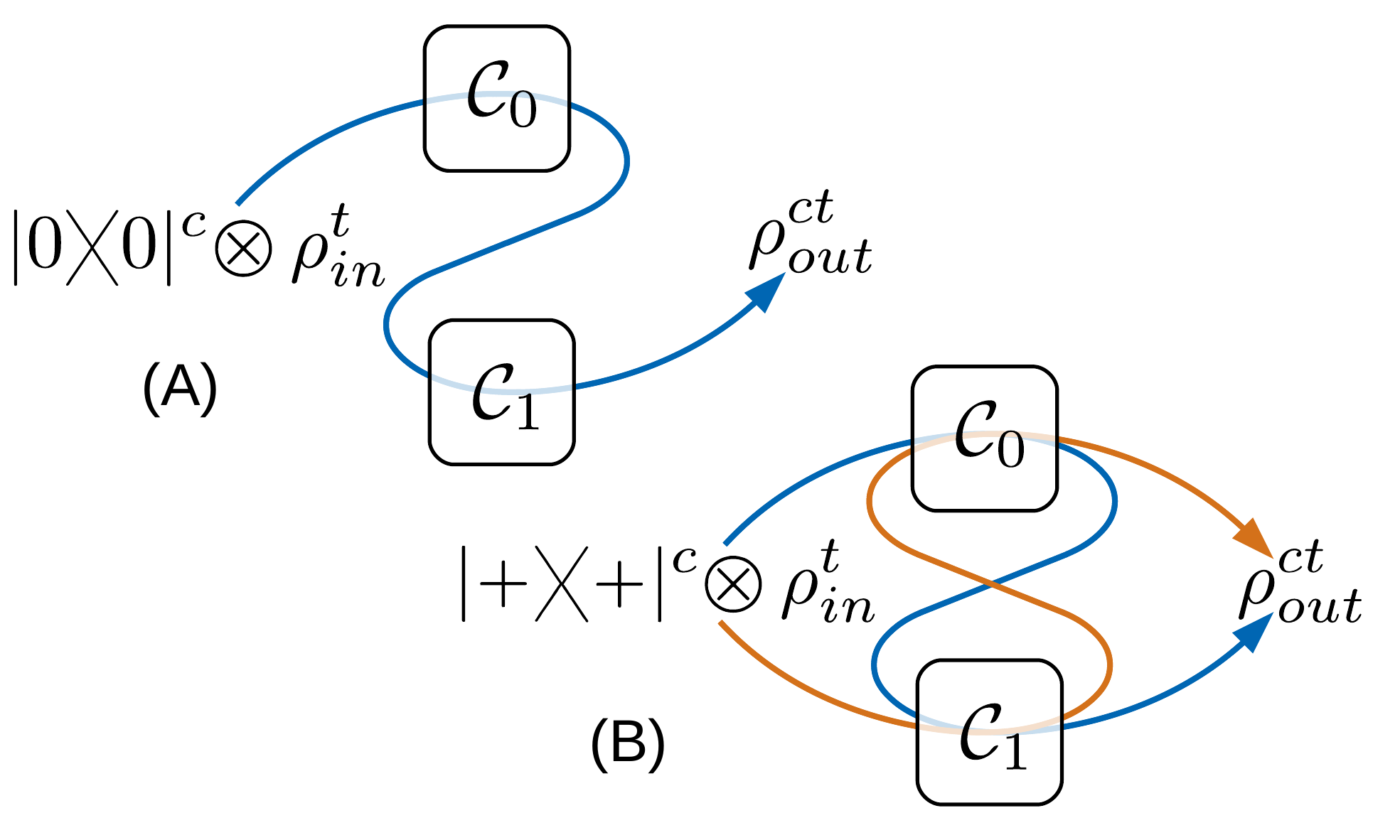}
\caption{Quantum circuit representing the quantum switch. Target system $\rho^t_{in}$ passes through $\C_0\circ\C_1$ and $\C_1\circ\C_0$ in a superposition determined by control qubit $c$. (A): the control is in the reduced state $\oprod{0}{0}^c$ and the target passes only through $\C_0\circ\C_1$. (B): the control is in the reduced state $\oprod{+}{+}^c$ and the target passes through a balanced superposition of $\C_0\circ\C_1$ and $\C_1\circ\C_0$.}
\label{fig:switch}
\end{figure}

A quantum switch $\C_0 \bowtie\C_1$ between channels $\C_0$ and $\C_1$ is a new channel that puts in a superposition two differently ordered compositions $\C_0\circ\C_1$ and $\C_1\circ\C_0$  (Figure \ref{fig:switch}). It acts on $\LI(\mathcal{H}^c)\otimes\LI(\mathcal{H}^t)$, where $c$ stands for control and $t$ for target. Unless stated otherwise, both of these Hilbert spaces are taken to be two-dimensional. The switch $\C_0\bowtie\C_1$ is a higher-order operation defined through its Kraus decomposition

\begin{equation}
V_{ij}=\oprod{0}{0}^c \otimes (K_i^{1}K_j^{0})^t + \oprod{1}{1}^c \otimes (K_j^{0}K_i^{1})^t \label{krausswitch},
\end{equation}
where $K _i ^0$ and $K_j ^1$ are the Kraus operators of $\C_0$ and $\C_1$ respectively. To get a heuristic picture, one may think about the target as a subsystem that passes through the channels, while the control subsystem in a generic state $\ket{\psi}^c=\alpha \ket{0}^c + \beta \ket{1}^c, |\alpha| ^2 + |\beta| ^2 =1,$  determines the order of passage. This is easy to comprehend in the case of unitary $U_0, U_1 \in \LI(\mathcal{H}^t)$ with the subsystems in pure states:
\begin{align}
(U_0 \bowtie U_1)\ket{\psi}^c\otimes\ket{\psi}^t = &\alpha\ket{0}^c \otimes U_0 U_1\ket{\psi}^t \\&+\beta\ket{1}^c \otimes U_1 U_0\ket{\psi}^t.\nonumber
\end{align}
One can see that, depending on the state of the control qubit, the order in which the target undergoes different operations is \textit{switched}.
If, for example, the state of the control qubit is $\ket{+}^c=\frac{1}{\sqrt{2}}(\ket{0}+\ket{1})^c$, then the quantum switch yields a balanced superposition of two orders.

Somewhat paradoxically, classical information can be transmitted though the quantum switch between two totally depolarizing channels. The Holevo capacity of a channel $\chi(\C)$ is defined as $\chi(\C) = \max_{\{p_a, \rho_a\}} I(A;B)_\nu$, where $\{\rho_a\}$ are the possible inputs of the channel with probabilities $p_a$ and $I(A;B)_\nu$ is the quantum mutual information calculated on the state $\nu = \sum_a p_a \oprod{a}{a}_A \otimes \C(\rho_a)_B$ \citep{holevo}. In terms of the von Neumann entropy $H$, if $\rho = \sum_a p_a \rho_a$, then
\begin{align}
    \chi(\C) &= \max_{\{p_a, \rho_a\}}\bigg[ H(\C(\rho))- \sum_a p_a H(\C(\rho_a)\bigg]. \label{holevoeq}
\end{align}
This maximum can be reached on no more than $N^2=4$ pure states \cite{Holevo2012}, where $N$ is the dimension of the target subsystem.

\begin{figure}[!htb]
\centering
\includegraphics[width=0.4\textwidth]{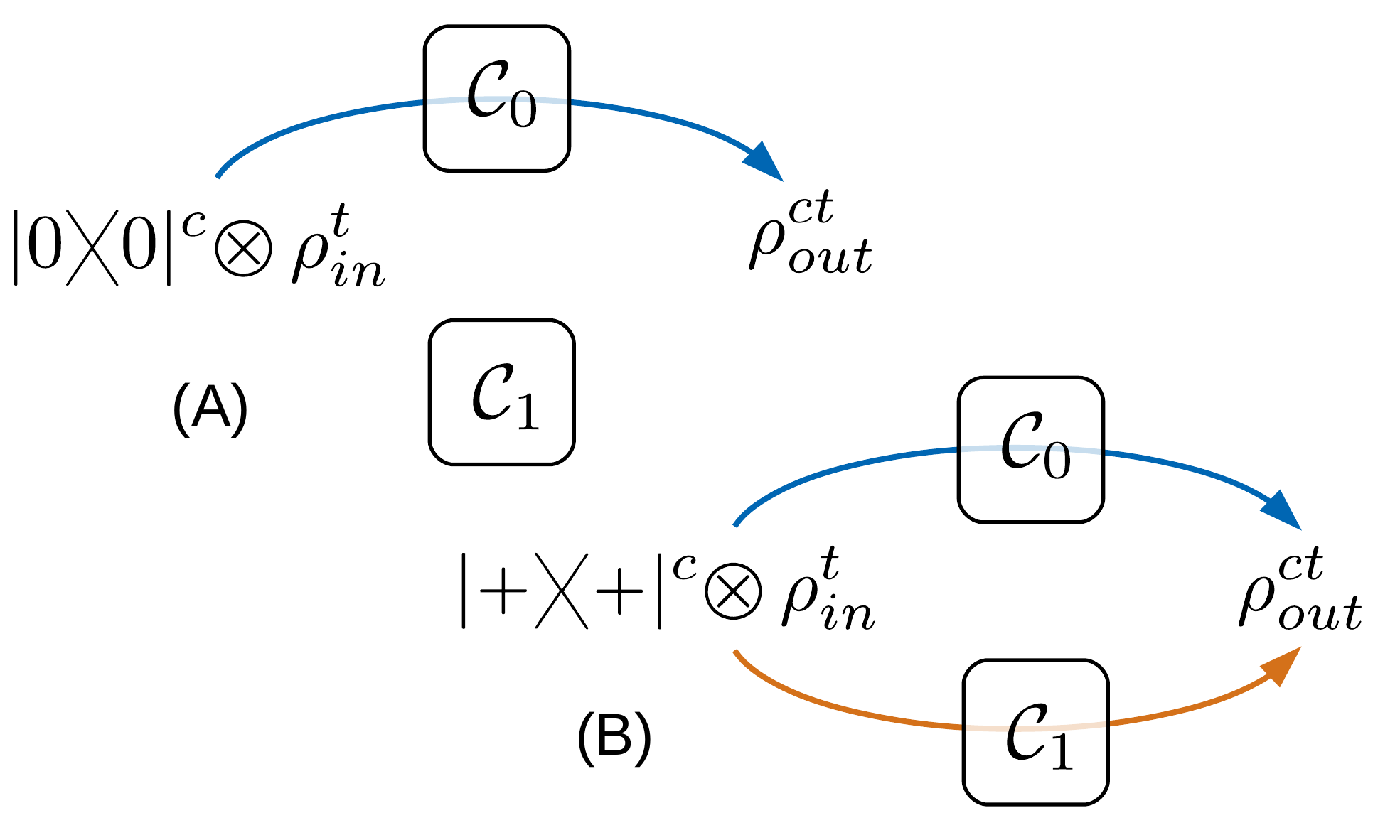}
\caption{Quantum circuit representing the one-pass superposition of $\C_0$ and $\C_1$. Target system $\rho^t_{in}$ passes through $\C_0$ and/or $\C_1$ in a superposition determined by control qubit ${\rho}^c$. (A): the control is in the reduced state $\oprod{0}{0}^c$ and the target passes only through $\C_0$. (B): the control is in the reduced state $\oprod{+}{+}^c$ and the target passes through a balanced superposition of $\C_0$ and $\C_1$.}
\label{fig:sup}
\end{figure}

The Holevo capacity of the quantum switch between two totally depolarizing channels acting on a qubit equals $\chi=-\frac{8}{3}-\frac{5}{8} \textrm{log}_2 \frac{5}{8} \approx 0.05$ \citep{Ebler}.
However, a similar advantage occurs if the channels are put in a superposition $C_0 \boxplus C_1$ (Figure \ref{fig:sup}). The one-pass superposition of $\C_0$ and $\C_1$ is defined through its Kraus operators
\begin{equation}
W_{ij}= \frac{1}{2} \oprod{0}{0}^c \otimes K_i^{0 t} + \frac{1}{2} \oprod{1}{1}^c \otimes K_j^{1 t}.
\end{equation}
As before, in the case of unitary $U_0$ and $U_1$ the definition can be simplified:
\begin{equation}
(U_0 \boxplus U_1)\ket{\psi}^c\otimes\ket{\psi}^t = \alpha\ket{0}^c \otimes U_0\ket{\psi}^t+\beta\ket{1}^c \otimes U_1\ket{\psi}^t.
\end{equation}

Starting from a well-known result in quantum communication~\cite{Gisin2005}, it has recently been shown that $\C_0 \boxplus \C_1$ has a greater Holevo capacity than $\C_0 \bowtie \C_1$ if $\C_0$ and $\C_1$ are totally depolarizing. A lower bound is $\chi (\C_0 \boxplus \C_1) \geq 0.16$ \citep{abbott}.

\section{Results}\label{section:results}

\begin{figure}[!htb]
\includegraphics[width=0.5\textwidth]{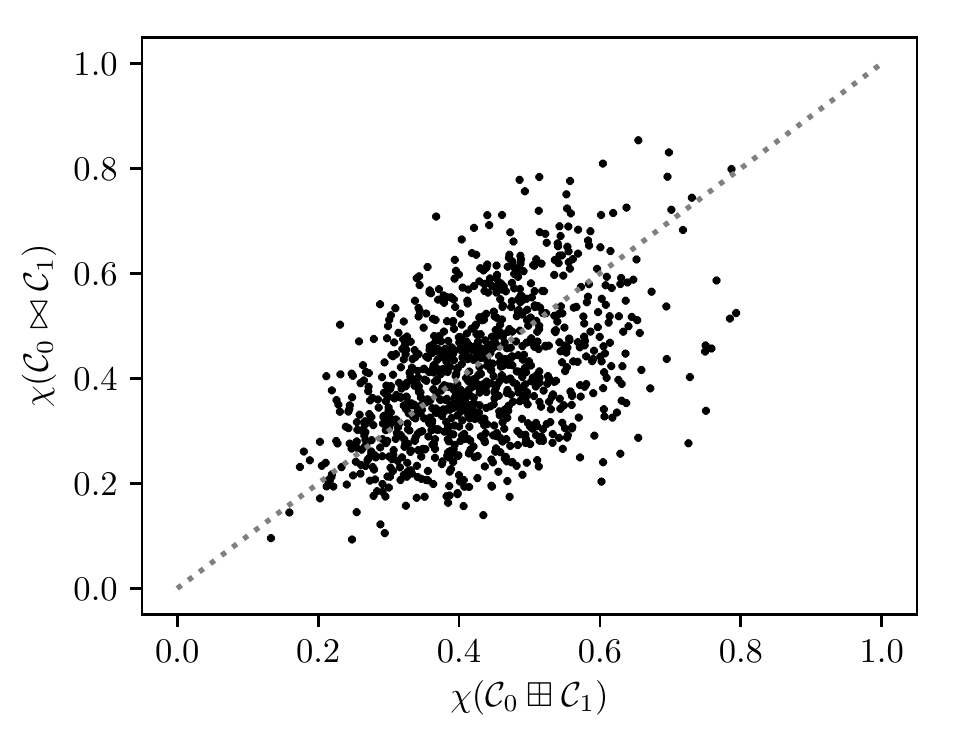}
\caption{\label{fig:switchvssup}Comparison of Holevo capacities of the quantum switch and the one-pass superposition. Each point corresponds to a couple of randomly generated quantum channels. Holevo capacities are estimated numerically.}
\end{figure}

The indefinite causal order provides an indisputable advantage in terms of Holevo capacity but this advantage is not systematic. To explore this situation, we randomly generate pairs of quantum channels $\C_0$ and $\C_1$ and numerically compute Holevo capacities of $\C_0 \bowtie \C_1$ and $\C_0 \boxplus \C_1$ for a control qubit in the reduced state $\oprod{+}{+}^c$.
Figure \ref{fig:switchvssup}, generated on two sets of 1000 channels each, shows the absence of any obvious correlation between $\chi(\C_0 \bowtie \C_1)$ and $\chi(\C_0 \boxplus \C_1)$. After three such runs, the average ratio $\chi(\C \bowtie \C) / \chi(\C_0 \boxplus \C_1)$ is stable around $0.9$, meaning that on average the one-pass superposition gives a slightly better advantage than the quantum switch. However, individual channels can exhibit vastly different behaviour.

\begin{figure}[!htb]
\includegraphics[width=0.5\textwidth]{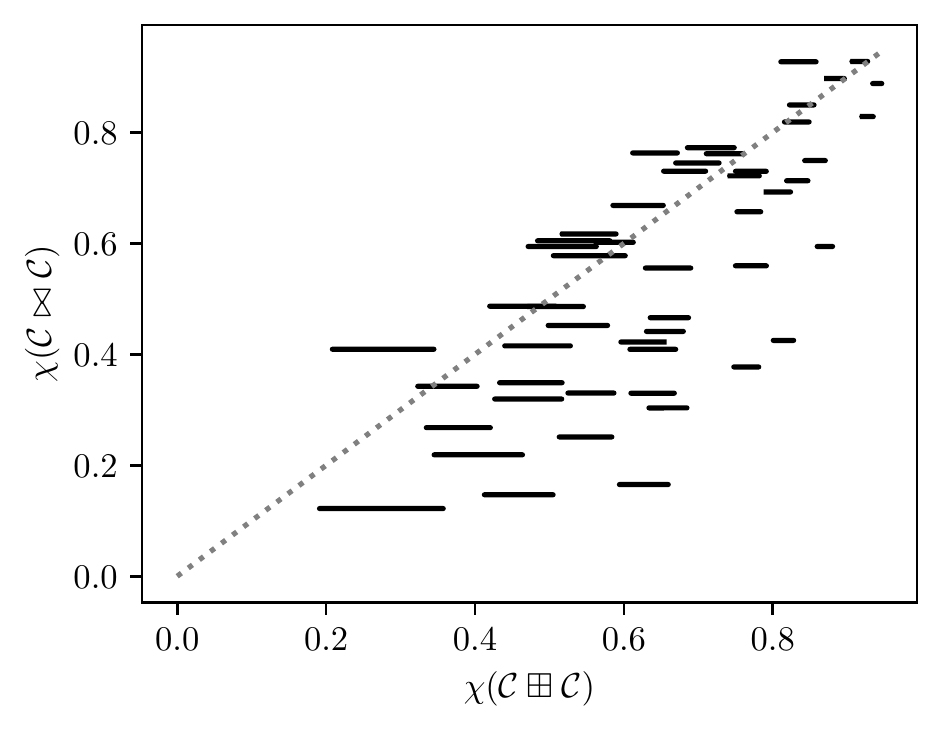}
\caption{\label{fig:switchvssupalt}Comparison of Holevo capacities of the quantum switch and the one-pass superposition for different implementations of $50$ randomly generated channels. Each line corresponds to $200$ different implementations of channel $\C$ due to the freedom in the Kraus decomposition. $\chi(\C \boxplus \C)$ depends on the implementation, while $\chi(\C \bowtie \C)$ is independent of the implementation.}
\end{figure}
\begin{figure}[!htb]
\includegraphics[width=0.5\textwidth]{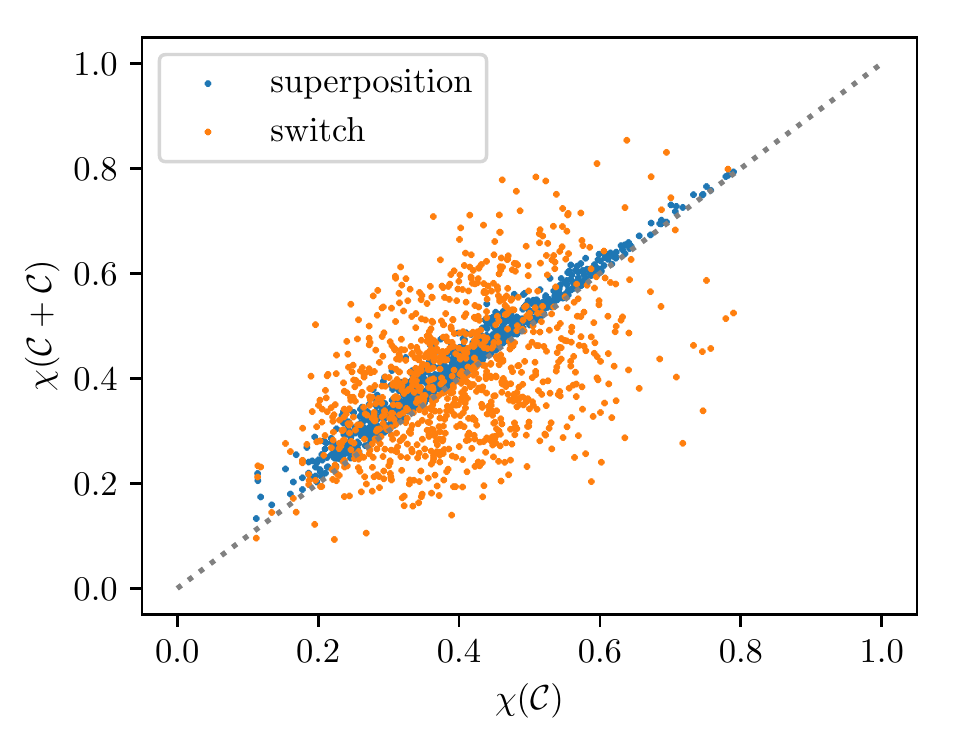}
\caption{\label{fig:holevo1}Comparison of Holevo capacities of self-switch and self-superposition of 1000 random channels. Holevo capacity always increases in the latter case but no such regularity exists in the former. The + sign stands alternatively for $\bowtie$ or $\boxplus$.}
\end{figure}
\begin{figure}[!htb]
\includegraphics[width=0.5\textwidth]{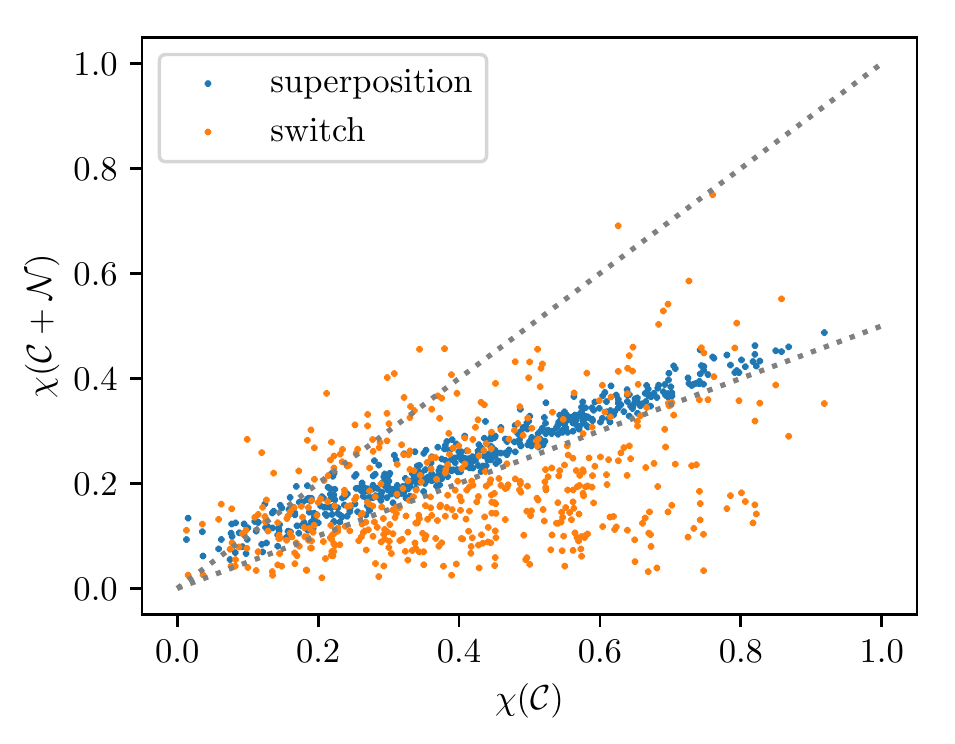}
\caption{\label{fig:holevo2}Comparison of Holevo capacities of the quantum switch and the one-pass superposition of a random channel $\C$ combined with a totally depolarizing channel $\N$. Each point corresponds to a random channel $\C$. Dotted line are linear functions with the slopes 1 and 1/2. The + sign stands alternatively for $\bowtie$ or $\boxplus$.}
\end{figure}

To study the combination of a channel with itself, we set $\C_0=\C_1=\C$. Figure \ref{fig:switchvssupalt} shows that $\chi(\C\boxplus\C)$ varies significantly as one applies \eqref{kraussfree} to change the Kraus decomposition of a channel, while $\chi(\C\bowtie\C)$ remains fixed.
Figure \ref{fig:holevo1} provides a comparison between the Holevo capacities of self-switch and self-superposition. The latter always increases channel capacity: $\chi(\C)<\chi(\C\boxplus\C)$ as we prove below, while this is not true for the former. If the number of subsystems is increased, e.g. in the case of the 3-switch \cite{Procopio2019}, the picture remains very similar to the one shown here.
To make sure that this effect is not only due to self-switching, we explore the Holevo capacity of a composition between a random channel and a totally depolarizing channel $\N$ with Kraus operators $\{\frac{\mathds{1}}{2}, \frac{\sigma_x}{2}, \frac{\sigma_y}{2}, \frac{\sigma_z}{2}\}$ (Fig. \ref{fig:holevo2}). All generated channels verify $\chi(\C)/2<\chi(\C \boxplus \N)$, whereas this inequality does not hold for the quantum switch.

\section{Methods}

A random channel $\C$ is obtained by generating a random set of Kraus operators. To get the latter, we generate a random Choi matrix. Let $X \in \LI (\HS)_A\otimes \LI (\HS)_B$ be a Ginibre complex random matrix and $Y=\text{Tr}_AXX^\dagg$. Then the following matrix is random Choi~\cite{Bruzda}:
\begin{equation}
    C = \left( \id_A \otimes \frac{1}{\sqrt{Y}} \right) XX^\dagg \left( \id_A \otimes \frac{1}{\sqrt{Y}} \right).
\end{equation}

Holevo capacity $\chi(\C)$ is computed by solving the optimization problem \eqref{holevoeq} using the basin-hopping method \cite{basinhopping}. States $\rho_a$ are parameterized as $\rho_a = \oprod{a}{a}$, $\ket{a}=(cos(\theta _a), sin(\theta _a)e^{i\phi _a})^T$, $(\phi _a, \theta _a)\in [0,\pi]\times[0,2\pi]$. Each step of the basin-hopping method consists of: a) random perturbation of the parameters, b) local minimization using the BFGS method, c) acceptance or rejection of the new parameters based on the minimized function value and the metropolis test \cite{Fletcher1987,metropolis}.

The basin-hopping method converges fast and is relatively easy to implement using common programming libraries. Since this method is heuristic by nature, we have compared its results with the tight bounds on Holevo capacity computed via a different procedure~\cite{Renner}. Both methods are in good agreement on all examples given in \cite{Renner}. To estimate the divergence, we have run the basin-hopping computation 20 times for each channel with different initial parameters and selected the highest estimated capacity. Standard deviation after 20 runs on a set of 1000 random channels is $<\sigma _\chi >=3.6\cdot 10^{-4}$ and $\max \sigma_\chi =1.1\cdot 10^{-2}$.

To check our results in the case of unitary channels only, we have also generated random unitary channels directly by a random set of unitary matrices $U_i$ and a random set of real coefficients $c_i$ constrained by $\sum_i c_i^2=1$. Then, Kraus operators $K_i=c_i U_i$. Any
such set of operators that verify $\sum_i K_i K_i^\dagger = \id$ defines a quantum channel. Using the Nelder-Mead method \cite{NeldMead65} with three free parameters, we have computed Holevo capacities and found that they are in full agreement with the results of the basin-hopping calculation.

\section{Discussion}\label{section:discussion}

To study the discrepancy between the capacities of the superposition and the quantum switch, note that the one-pass superposition acts as:
\begin{align}(\C\boxplus\C)(\rho) = \frac{1}{2} \left(\begin{array}{@{}c|c@{}}
\C(\rho) & \sum_{i,j}K_i \rho K_j^\dagger \\
\hline
\sum_{i,j}K_i \rho K_j^\dagger&\C(\rho)
\end{array}\right)
\label{csupc},
\end{align}
where $\{K_i\} $ are the Kraus operators of $\C$ and the fixed control subsystem is omitted for brevity. To prove $\chi(\C\boxplus\C) \geq \chi(\C)$, define a channel $\mathcal{P}$ acting on $\LI(\mathcal{H}^c)\otimes\LI(\mathcal{H}^t)$ with Kraus operators $\{P_m\} = \{\oprod{m}{m}^c \otimes \mathbb{1}^t\}$. Since $\sum_m P_m P_m^\dagger = \mathbb{1}^{ct}$, $\mathcal{P}$ is trace-preserving. It acts as:\begin{align}\mathcal{P}\circ(\C\boxplus\C)(\rho) =\C'(\rho)=\frac{1}{2} \left(\begin{array}{@{}c|c@{}}
\C(\rho) & 0\\
\hline
0&\C(\rho)
\end{array}\right).
\end{align}

If the spectrum of $C(\rho)$ is $\{\eta_i\}$, then the spectrum of $C'(\rho)$ is $\{\frac{\eta_i}{2}\}$, each eigenvalue having multiplicity two. Hence:
\begin{align}
H(C'(\rho))&=-\sum^8_{i=1}\frac{\eta_i}{2}\log\frac{\eta_i}{2} \nonumber\\
&= -\frac{1}{2}\sum^8_{i=1}\eta_i( \log(\eta_i) -1) \nonumber\\
&= \sum^4_{i=1} \eta_i \log(\eta_i)+ \sum^4_{i=1}\eta_i \nonumber\\
&= H(C(\rho))+1.
\end{align}

It now follows from \eqref{holevoeq} that:
\begin{align}
    \chi(\C') &= \max_{\{p_i, \rho_i\}}\bigg[ H(\C'(\rho))- \sum_i p_i H(\C'(\rho_i))\bigg]\nonumber\\
    &= \max_{\{p_i, \rho_i\}}\bigg[ H(\C(\rho)) -\sum_i p_i H(\C(\rho_i))\bigg]\nonumber\\
    &=  \chi(\C),
\end{align}
i.e. $\chi(\mathcal{P}\circ(\C\boxplus\C)) = \chi(\C)$. By the data processing inequality, a channel can only lose information between the input and the output:
$\chi(\C) = \chi (\mathcal{P}\circ(\C\boxplus\C)) \leq \chi (\C\boxplus\C)$.\qed

The quantum switch acts as:
\begin{multline}(\C \bowtie\C)(\rho)  = \\
\frac{1}{2} \left(\begin{array}{@{}c|c@{}}
\C(\C(\rho)) & \sum_{i,j} K_jK_i\rho K_j^\dagger K_i^\dagger \\
\hline
\sum_{i,j} K_iK_j\rho K_i^\dagger K_j^\dagger  & \C(\C(\rho))
\end{array}\right)\label{cswitchc}.
\end{multline}
Unlike with quantum superposition, no simple relationship between Holevo capacities is available in this case. When the quantum switch lowers communication capacity of $\C$, this phenomenon has a likely origin in the loss of information in diagonal terms $\C \circ \C$. On the contrary, the advantage of the quantum switch over the superposition, when it occurs, comes from non-diagonal terms. Heuristically, the non-diagonal terms in \eqref{cswitchc} are more versatile than the non-diagonal terms in \eqref{csupc}, explaining the behaviour of the switch in Figure \ref{fig:holevo1}.

Note that the self-switch $\C \bowtie \C$ is not the same channel as the superposition $(\C\circ\C) \boxplus (\C\circ\C)$. If $\C$ has Kraus operators $\{K_i\}$, then the Kraus operators of $\C\circ\C$ are $\{K_iK_j\}$. Inserting this in \eqref{csupc}, one obtains:
\begin{align}((\C\circ\C) \boxplus (\C\circ\C))(\rho) & = \frac{1}{2} \left(\begin{array}{@{}c|c@{}}
\C(\C(\rho)) & \C(\C(\rho)) \\
\hline
\C(\C(\rho))  & \C(\C(\rho))
\end{array}\right).\label{ccsupcc}
\end{align}
This expression is different from \eqref{cswitchc} if $\{K_i\}$ do not commute. The latter factor happens to have an even deeper significance: our study of the non-diagonal terms provides evidence for the hypothesis, inspired by \cite{Ebler}, that a greater advantage occurs when the Kraus operators do not commute. To observe this effect, define:
\begin{align}
    Q(\C) &= \sum_{i,j}\textrm{Tr}\left([K_i,K_j][K_i,K_j]^\dagg\right)\label{Q}\\
    &=4 - 2\textrm{Tr}\sum_{i,j} K_i K_j K_i^\dagg K_j^\dagg \nonumber.
\end{align}
$Q(\C)$ is independent of the Kraus decomposition of $\C$.
Indeed, note that two Kraus decompositions $\{E_i\}$ and $\{F_i\}$ describing the same channel are related by \eqref{kraussfree}:
\begin{align*}
    \sum_{i,j}E_i E_j E^\dagg_i E^\dagg_j
    & = \sum_{i,j}\sum_{k,l,m,n}u_{ik}u_{im}^\dagg u_{jl} u_{jn}^\dagg F_k F_l F_m^\dagg F_n^\dagg \\
    & = \sum_{k,l,m,n}\delta_{km} \delta_{ln} F_k F_l F_m^\dagg F_n^\dagg \nonumber \\
    & = \sum_{i,j}F_i F_j F_i^\dagg F_j^\dagg \nonumber.
\end{align*}
Hence $Q(\{E_i\}) = Q(\{F_i\})$.\qed

\begin{figure}
\centering
\includegraphics[width=0.5\textwidth]{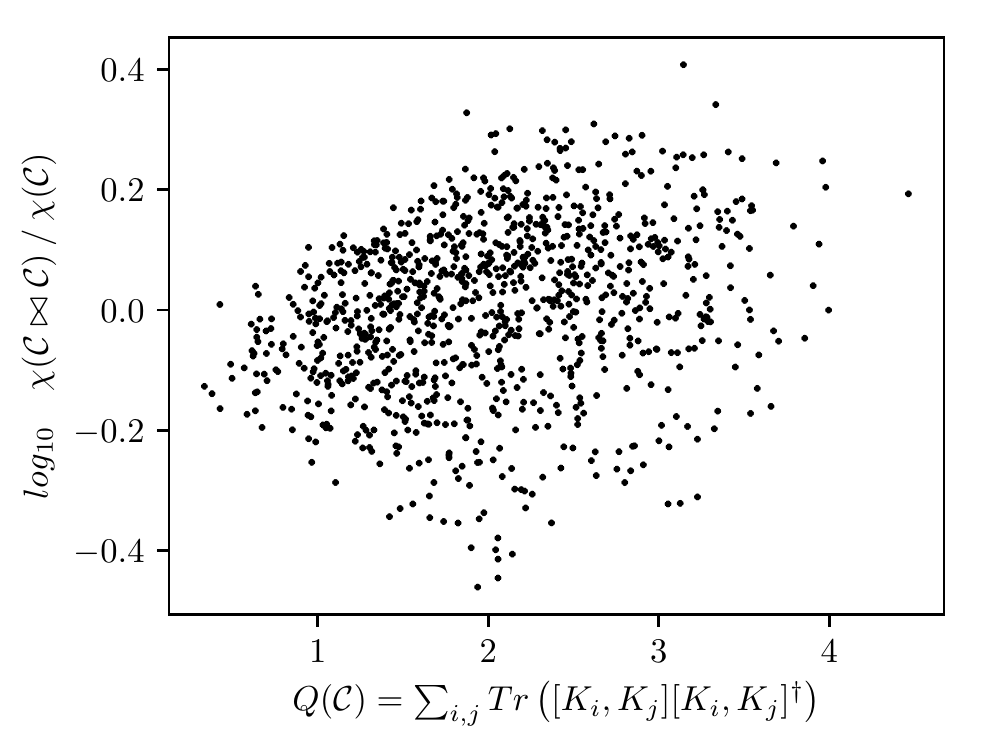}
\caption{Relative advantage of the quantum switch $\chi(\C \bowtie\C)/\chi(\C)$ as a function of $Q(\C)$, degree of commutativity of the Kraus decomposition, on a set of $1000$ random channels.
While the quantum switch can gain or lose communication capacity depending on $\C$, it does so more often at higher values of $Q(\C)$.
}
\label{fig:commutators1}
\end{figure}

\begin{figure}
\centering
\includegraphics[width=0.5\textwidth]{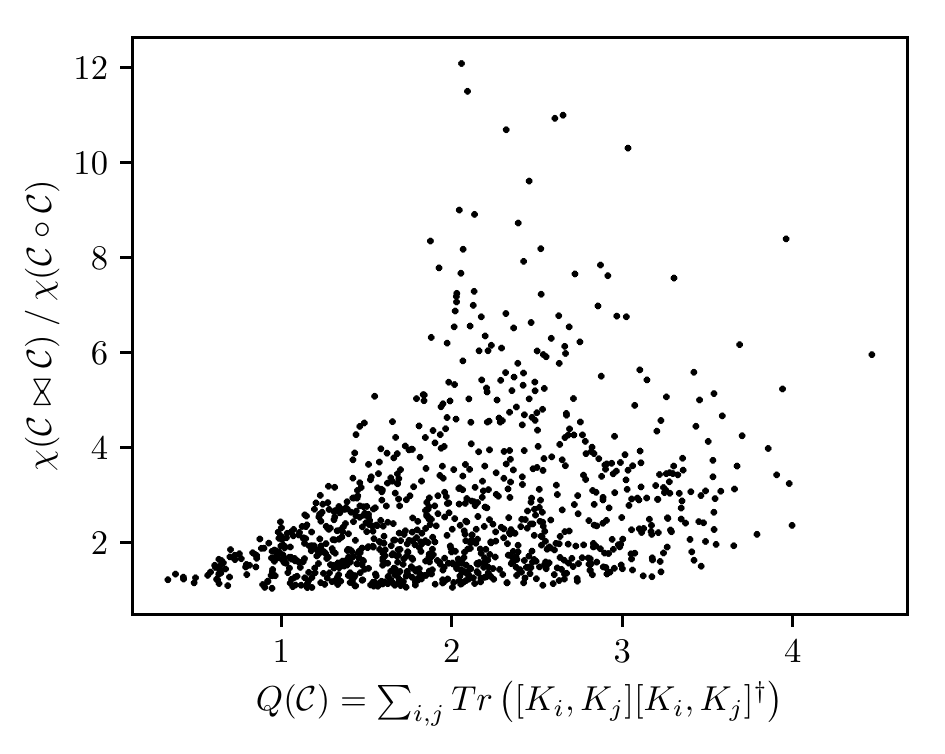}
\caption{Advantage $\chi(\C \bowtie \C)/\chi(\C\circ \C)$ as a function of $Q(\C)$, the degree of commutativity of the Kraus decomposition. If the Kraus operators nearly commute, then the Holevo capacities are similar. If they do not commute, then the effect of the indefinite causal order gets stronger.}
\label{fig:commutators2}
\end{figure}

Figures \ref{fig:commutators1} and \ref{fig:commutators2}, generated on a set of $1000$ random channels, show that the spread of $\chi(\C \bowtie \C)/\chi(\C)$ increases with $Q$. If $\{K_i\}$ almost commute and $Q$ is small, then the effect of the indefinite causal order is also small: the Holevo capacity of the quantum switch is close to the Holevo capacity of the superposition. On the contrary, when $Q$ is high, the effect of the indefinite causal order is strong: the Holevo capacity of the quantum switch is dominated by this non-commutativity and the non-diagonal terms of \eqref{cswitchc} are larger than the diagonal ones. Our simulation shows that the same conclusion is also valid for the 3-switch \cite{Procopio2019}.

\section{Conclusion}

Both the coherent control of quantum channels and the indefinite causal order between them were shown to be resources for communicating through noisy channels. While the notion of controlling quantum channels via a quantum subsystem is a useful heuristic, its rigorous mathematical formulation shows that the action of observers involved in such control and measurement operations is hardly ever local in time and space \cite{Oreshkov2018,Brukner2019}. Nonetheless, as mentioned above, experimental setups have been realized for two or three subsystems combined via the quantum switch. These empirically available setups call for a better theoretical characterization of their workings.

On the one hand, it has been argued that the communication advantage found with the quantum switch ``is also present'' in the scenario involving one-pass superposition~\cite{abbott}. On the other, this advantage was hailed as a ``striking \ldots new paradigm of Shannon theory''~\cite{Ebler}. Our analysis via numeric simulations shows that the two resources provide the same communication gain only in very rare cases. In a more general setting, simulated by choosing either generic or unitary channels at random, the conceptual origin and the quantitative measure of the advantage differ. This is because the quantum switch exhibits an intricate interplay between two non-classical factors, both of which are capable of providing a communication advantage: quantum superposition and non-commutativity. What is more, the quantum switch, but not the one-pass superposition, can also induce a loss in communication capacity because it involves a composition of channels. When the switch is treated as a resource, it remains a genuine combination of different contributing properties and should not be reduced to any one of them.  These results motivate further theoretical research on an important problem of quantum communication: achieving precise understanding of the interrelations between non-classical resources that provide an advantage in communication via noisy channels.

%

\end{document}